\documentclass[aps,twocolumn,prl,showpacs,floatfix,superscriptaddress,reprint]{revtex4-1}
\usepackage{amsmath,fixmath}
\usepackage{graphicx}
\usepackage{dcolumn}
\usepackage{amssymb}
\usepackage{color}
\usepackage{natbib}
\usepackage{hyperref}

\begin{document}
\textbf{Comment on ``Evidence of Non-Mean-Field-Like Low-Temperature
Behavior in the Edwards-Anderson Spin-Glass Model''}\newline
Ref.~\cite{yucesoy:12} compares the low-temperature
phase of the $3D$  Edwards-Anderson model (EA) to 
the Sherrington-Kirkpatrick model (SK), studying
the overlap distributions  $P_\mathcal{J}(q)$ and concluding
that the two models behave differently. A similar
analysis using state-of-the-art, larger data sets for EA
(generated with Janus~\cite{janus:12b} in
\cite{janus:10}) and for SK (from~\cite{aspelmeier:08})
leads to a very clear interpretation of the results of
\cite{yucesoy:12}, showing that EA behaves as predicted by
the replica symmetry breaking (RSB) theory.

Ref.~\cite{yucesoy:12} studies $\Delta(\kappa,q_0)$, probability of
finding in $P_\mathcal{J}(q)$ a peak greater than $\kappa$ for $q<q_0$.
In a RSB system, $\lim_{N\to\infty} \Delta(\kappa, q_0)=1$.  Fig.~5
of~\cite{yucesoy:12} shows that, at fixed $q_0$ and at the same
$T/T_\text{c}$,
$\Delta$ grows for SK, but seems to reach a plateau for EA. In the inset
of Fig.~1 we show that, considering larger systems ($N\leq32^3$ as
opposed to $N\leq12^3$ of~\cite{yucesoy:12}), $\Delta$ clearly grows
with $N$ also for EA.  We use the same value of $q_0$ as
in~\cite{yucesoy:12} and $T=0.703$.  Even this simple analysis is
sufficient, when using state-of-the-art lattice sizes, to show that
$\Delta$ has the same qualitative behavior in both models.

Still, the choice of comparing data for different models at the same 
$T/T_\text{c}$ and $N$ does not have a strong basis.  Indeed, according to the mean-field
picture, the fluctuations of the $P_\mathcal{J}(q)$ are ruled by the shape of
the averaged $P(q)$~\cite{parisi:93}, so it is more appropriate 
to select $T$ such that $P(q)$ is similar for EA and SK. Now, 
it is universally accepted that the peak at $q=q_\text{EA}$   
in $P(q)$ grows with $N$ more slowly for EA, so the simplest
assumption that all the individual peaks for $q<q_\text{EA}$ scale at
the same rate would already explain the results reported in~\cite{yucesoy:12}.

According to RSB theory, in the large-$N$ limit $P_\mathcal{J}(q)
= \sum_{\gamma} W_\gamma \delta(q-q_\gamma)$. Let 
us assume that for large but finite $N$ the weight distribution is
unchanged, but the delta functions are smoothed to a finite
height $H(N)$~\cite{janus:11}. The self-averaging
peak at $q=q_\text{EA}$ will also be smoothed, 
so we can estimate $H(N)\sim P(q_\text{EA})$.  $\Delta(\kappa, q_0)$ is the
probability of finding a peak with weight $W_\gamma > \kappa/H(N)$,
which, for small $q_0$, is $\Delta(\kappa,q_0) \sim
[\kappa/H(N)]^{-I(q_0)}\sim \bigl(P(q_\text{EA})/\kappa\bigr)^{I(q_0)}$, where
$I(q_0)=\mathbb{P}(|q|<q_0)$~\cite{parisi:93}.

\begin{figure}[b]
\centering
\includegraphics[width=\columnwidth,angle=270]{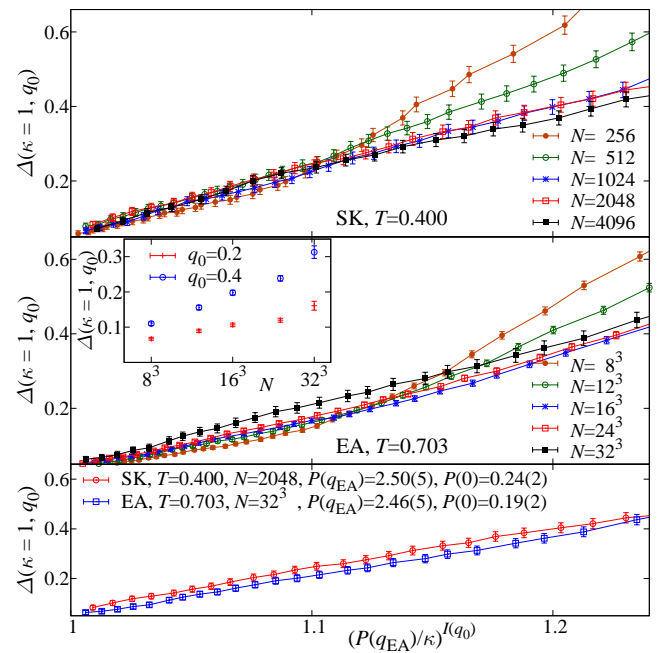}
\caption{(color online) $\Delta(\kappa,q_0)$ against
  $[P(q_\text{EA})/\kappa]^{I(q_0)}$ for SK (\emph{top}) and EA
  (\emph{middle}).  \emph{Inset:} $\Delta(\kappa,q_0)$ for fixed $q_0$
  for the EA model. \emph{Bottom:} comparison of the EA and SK models
  for similar values of $P(q_\text{EA})$.}
\end{figure}

We show $\Delta$ at $T=0.4$ for SK (top) and at $T=0.703$
for EA (middle), where the temperatures are such that $P(0)$ are
very similar (for the largest systems, $q_0$ ranges from $0.02$ to $0.44$).
The curves show universal scaling for large
$N$. The bottom panel compares $\Delta$  for SK and EA 
using similar effective sizes. 

In short, the simple assumption that peaks for all 
values of $q$ scale at the same rate is consistent
with the numerical data and explains the slower growth
of $\Delta$ with $N$ for EA. Therefore, 
contrary to the claims in~\cite{yucesoy:12}, 
we find no quantitative difference between EA and SK, 
as long as one is careful when comparing non-universal quantities 
and uses state-of-the-art system sizes.

We have been supported through research contracts nos.\ 247328 (ERC);
FIS2012-35719-C02-01 and FIS2010-16587 (MICINN); and GR10158 (Junta de
Extremadura). We thank the Janus Collaboration for granting us use of
the EA data.

\vspace*{\baselineskip}

\noindent A.~Billoire$^1$, L.A.~Fernandez$^2$, A.~Maiorano$^3$,
E.~Mari\-nari$^3$, V.~Martin-Mayor$^2$, G.~Parisi$^3$,
F.~Ricci-Tersenghi$^3$, J.J.~Ruiz-Lorenzo$^4$, D.~Yllanes$^3$.\\
\noindent $^1$IPhT, CEA Saclay, 91191 Gif-sur-Yvette, France;\\
$^2$Dep.~F{\'\i}sica Te\'orica I, UCM, 28040 Madrid, Spain;\\
$^3$Dip.~Fisica, La Sapienza Universit\`a di Roma, 00185 Roma, Italy;\\
$^4$Dep.~F{\'\i}sica, Univ.~Extremadura, 06071 Badajoz, Spain
\end{document}